\documentstyle[aps,prc]{revtex}
\begin{document}
\draft
\title{
Effects of the difference between the charge and matter deformations
 on fusion reactions of unstable nuclei
}
\author{Tamanna Rumin
\thanks{Electronic address: rumin@nucl.phys.tohoku.ac.jp}
and
Noboru Takigawa,
\thanks{Electronic address: takigawa@nucl.phys.tohoku.ac.jp}
}
\address{Department of Physics, Tohoku University, 
Sendai 980-8578, Japan \\
}
\date{\today}

\maketitle

\begin{abstract}

Relativistic mean field calculations suggest that 
the charge and matter deformations significantly differ in some of the 
unstable neutron and proton rich nuclei. We discuss the effects of 
the difference on the fusion reactions induced by them at energies near 
and below the Coulomb barrier by taking 
the $^{19,25,37}$Na + $^{208}$Pb reactions as examples.  
We also discuss whether one can probe 
the difference by the so called fusion barrier distribution analysis. 

\end{abstract}


\section{Introduction}

Secondary beam experiments are opening up 
new physics to be studied for a variety of unstable nuclei 
across the beta-stability line in the nuclear chart. 
One of the basic questions is to study their charge and matter 
deformations. 
We are especially interested 
in the difference between the charge and matter deformations in each 
unstable nucleus. 
One cannot, however, apply the standard techniques such as the 
gamma ray spectroscopy which have been used for stable nuclei. 
A possibility is to use heavy-ion fusion reactions at energies below or 
near the Coulomb barrier, which have recently been established to be 
very sensitive to the shape of the target and/or projectile nuclei 
\cite{bt98,dhrs98}. An alternative is to predict the charge and matter 
deformations 
theoretically based on the non-relativistic Hartree Fock or 
the relativistic mean field calculations with various 
refinements such as including the pairing correlations. 

In this paper, we take Na isotopes as an example to anticipate the 
variation of the charge and matter deformations along an isotope chain 
and discuss whether the significant difference between them 
expected for some isotopes can be probed through 
the analysis of the fusion reactions induced by them
by assuming $^{208}$Pb as the target nucleus. 
We use the relativistic mean field calculations to 
assess the charge and matter deformations for each Na isotope. 
We then calculate the 
fusion cross section $\sigma(E)$ based on the 
coupled-channels formalism assuming thus obtained values of 
deformations and taking the ground state rotational band into account. 
We then calculate the associated so called fusion barrier distribution 
$\frac{d^2(E\sigma(E))}{dE^2}$ \cite{rss91}. 

The paper is organized as follows. In sect.2 we describe the formalism 
of our coupled-channels calculations, i.e. the simplified 
framework based on the no-Coriolis 
approximation, somewhat in detail, because 
the formalism is not well known for the case of fusion of 
a deformed odd mass projectile 
which we are going to discuss in this paper. 
In sect. 3, we present the results of the relativistic mean field 
calculations and of the coupled-channels analyses. The paper is 
summarized in sect.4. 

\section{Simplified coupled-channels formalism for the fusion 
of a deformed odd-mass nucleus}

We discuss in the next section the fusion of 
deformed odd mass Na isotopes with $^{208}$Pb target. We calculate the 
fusion cross section by taking the excitation of the ground state 
rotational band of the Na projectile into account 
with the coupled-channels formalism in the no-Coriolis 
approximation. In this section, we describe the major aspects of the 
method. 

We denote the coordinates of the relative motion between the 
projectile and target nuclei 
and of the rotational motion of the projectile, 
i.e. a Na isotope, by ${\vec R}$ and $\hat{\xi}$, 
respectively. 
We ignore the intrinsic excitation of $^{208}$Pb. The total Hamiltonian 
is then given by
\begin{eqnarray}
H(\vec R,\xi)=\hat T_R+V_0(R)+H_0(\xi)+V(\vec R,\xi)
\label{totalh}
\end{eqnarray}
with 
\begin{eqnarray*}
\hat T_R=-\frac{\hbar^2}{2\mu}\frac{1}{R^2}\frac{d}{d 
R}R^2\frac{d}{dR}+\frac{\hbar^2}{2 \mu R^2}\hat L^2
\end{eqnarray*}
$V_0(R)$ is the sum of monopole nuclear 
and Coulomb interactions between the projectile and target nuclei. 
We call it the bare interaction.
$H_0(\xi)$ is the intrinsic Hamiltonian of each Na 
projectile, whose eigenfunctions and the eigenvalues are 
denoted as $\varphi_{Im_I}^K(\xi)$ and $\epsilon_{IK}$, 
\begin{eqnarray}
H_0(\xi)\varphi_{Im_I}^K(\xi)=\epsilon_{IK}\varphi_{Im_I}^K(\xi)
\label{intse}
\end{eqnarray}
We assume that each deformed Na isotope can be well described 
by the strong coupling model \cite{bmII} for the rotational excitation, 
where the wave function is given by 
\begin{eqnarray}
\varphi_{Im_I}^K(\xi)=\sqrt{\frac{2I+1}{16\pi^2}}\Big\{D_{m_IK}^I(\Omega)\chi_K+(-1)^{I+K}D_{m_I-K}^I(\Omega)\chi_{-K}\Big\}
\label{iwf}
\end{eqnarray}
On the r.h.s. of eq.(\ref{iwf}), 
we used $\Omega$ for $\xi$ 
in order to explicitly indicate the physical meaning as the Euler 
angles. 
$\chi_K$ is the intrinsic wave function. 
We assume the deformation to be axially symmetric, and K is the 
component of the angular momentum of the 
rotational motion along the symmetry axis. 

The $V(\vec R,\xi)$ in eq.(\ref{totalh}) represents the interaction 
between the projectile and 
target nuclei after the monopole part has been subtracted. 
We assume the same form of coupling Hamiltonian 
as that is familiar in coupled-channels 
calculations for the fusion reactions of even-even nuclei. 
It consists of the nuclear and Coulomb parts, and is expressed as 
\begin{eqnarray}
V(\vec R,\xi)=f^N(R)\sum_{\lambda } Y_\lambda(\hat R)\cdot 
T_\lambda(\xi)+\sum_{\lambda\mu}f^C_\lambda(R) Y_{\lambda\mu}(\hat R)
T^\star_{\lambda\mu}(\xi)
\label{cint}
\end{eqnarray}
in terms of the spherical harmonics for the relative motion 
and the tensor operators ${\hat {T}}_{\lambda \mu}$ 
which cause the rotational excitation of the Na projectile. 
We further assume the latter to be given by 
\begin{eqnarray}
T_{\lambda\mu}(\xi)=\sum_\nu a_{\lambda \nu}{\cal D}^\lambda_{\mu \nu}(\Omega)
=a_{\lambda 0}{\cal D}^\lambda_{\mu 0}(\Omega)=
\beta_\lambda {\cal D}^\lambda_{\mu 0}(\Omega)
\label{ctensorin}
\end{eqnarray}
where $a_{\lambda \nu}$ are the deformation parameters in the body 
fixed frame \cite{bmII}. 
We have kept only 
the $\nu=0$ term to be consistent 
with the assumption of the axially symmetric deformation. 
The deformation parameter $\beta_\lambda$ is identified with 
the nuclear and the charge deformation parameters, $\beta^N_\lambda$ 
and $\beta^C_\lambda$, 
in the nuclear and Coulomb coupling terms, respectively. 
The $f^N(R)$ and $f^C_\lambda(R)$ are the nuclear and Coulomb coupling 
form factors, which we assume to be, 
\begin{eqnarray*}
f^N(R) =-R_P \frac{d}{dR}[\frac{-V_0}{1+exp[R-R_0]/a_0}], 
\hskip 1cm
f^C_\lambda(R) =\frac{3Z_PZ_Te^2}{2\lambda+1}\frac{R_P^{\lambda}}
{R^{\lambda+1}} 
\label{ff}
\end{eqnarray*}
with $R_0=R_P+R_T$, $R_P$ and $R_T$ being the nuclear radii of the 
projectile and target. 

As usual, we introduce the channel wave functions by 
\begin{eqnarray}
\phi_{IKL}^{JM}(\hat R,\xi)=\sum_{m_Lm_I}(Lm_LIm_I|JM)Y_{Lm_L}(\hat R)\varphi_{Im_I}^K(\xi)
\end{eqnarray}
and expand the total wave function for a fixed total angular momentum $J$ 
and its z component $M$ as, 
\begin{eqnarray}
\Psi^{JM}(\vec R, \xi)=\sum_{IKL}\frac{1}{R}\Phi_{IKL}^J (R)\phi_{IKL}^{JM}
(\hat R,\xi)
\end{eqnarray} 
It is then straight-forward to obtain the following coupled 2nd order 
differential equations which 
determine the radial wave functions of the relative motion $\Phi_{IKL}^J (R)$, 
\begin{eqnarray}
\Big[-\frac{\hbar^2}{2\mu }\frac{d^2}{dR^2}+V_0(R)+\frac{\hbar^2}{2\mu 
R^2}L(L+1)+\epsilon_{IK}-E\Big]\Phi_{IKL}^J (R)+\sum_{\lambda I^\prime K^\prime L^\prime}V_{IKL,I^\prime K^\prime 
L^\prime}(R)\Phi_{I^\prime K^\prime L^\prime}^J (R)=0
\label{GCC}
\end{eqnarray}
where the coupling matrix elements are given by
\begin{small}
\begin{eqnarray}
V_{IKL,I^\prime K^\prime L^\prime}(R)&=&\Big[f^N(R) \beta^N_\lambda+ 
f^C_\lambda(R) \beta^C_\lambda\Big]
(-1)^{L+J+L^\prime+I^\prime}[\frac{1+(-1)^\lambda}{2}] 
\sqrt{\frac{(2I^\prime+1)(2L+1)(2\lambda+1)(2L^\prime+1)}{4\pi}}\nonumber \\
&&\left(\begin{array}{ccc}L & \lambda & L^\prime \\
0 & 0& 0
\end{array}\right)
\left\{\begin{array}{ccc}L & I & J \\
I^\prime & L^\prime & \lambda
\end{array}\right\} 
(\lambda 0 I^\prime K|IK)\delta_{KK^\prime}
\label{cmat}
\end{eqnarray}
\end{small}
In obtaining eq.(\ref{GCC}) with eq.(\ref{cmat}), we used the explicit 
expression of the tensor operator $T_{\lambda \mu}$ 
given by eq.(\ref{ctensorin}), which leads to 
the selection rule $\delta_{KK^\prime}$.

The full coupled-channels equations (\ref{GCC}) become fairly 
intricate if many channels are included. In this paper, we use the 
no-Coriolis approximation, in other words the rotating frame 
approximation, 
which considerably reduces the dimension of the coupled equations 
\cite{ti}.  
The process to calculate the fusion cross section then 
consists of three major steps. The first is to 
replace the channel dependent centrifugal potential energy in 
eq.(\ref{GCC}) 
by that in the entrance channel, 
\begin{eqnarray}
\frac{L(L+1)\hbar^2}{2\mu R^2}\approx \frac{L_i(L_i+1)\hbar^2}{2\mu R^2}
\end{eqnarray} 
where $L_i$ is the initial orbital angular momentum. 

The second step is to perform the unitary transformation, 
\begin{eqnarray}
\tilde \Phi_{I \tilde K }^{JK}(R)=\sum_L\sqrt{\frac{2L+1}{2J+1}}(L0I\tilde K|J\tilde K)\Phi_{I 
K L}^J (R)
\label{Twf}
\end{eqnarray}
which transforms the original coupled-channels equations eq.(\ref{GCC}) 
into 
\begin{eqnarray}
\Big[-\frac{\hbar^2}{2\mu 
}\frac{d^2}{dR^2}+V_0(R)+\frac{L_i(L_i+1)\hbar^2}{2\mu R^2}
+\epsilon_{IK}-E\Big]\tilde \Phi_{I\tilde K}^{JK}(R)
+\sum_{\lambda I^\prime }V^\prime_{I\tilde K,I^\prime \tilde K}(R)
\tilde \Phi_{I^\prime \tilde K }^{JK} (R)=0
\label{RCC}
\end{eqnarray}
with 
\begin{eqnarray}
V^\prime_{I\tilde K,I^\prime \tilde K}(R)=(-1)^{I+2\lambda-I^\prime}
\Big[f^N(R) \beta^N_\lambda+ f^C_\lambda(R) 
\beta^C_\lambda\Big][\frac{1+(-1)^\lambda}{2}]
\sqrt{\frac{(2\lambda+1)(2I^\prime+1)}{4\pi (2I+1)}} 
(\lambda 0 I^\prime K|IK) 
(\lambda 0 I^\prime -\tilde K |I -\tilde K)
\label{ccmd}
\end{eqnarray}
The $\tilde K$ is the projection of the intrinsic spin 
of the Na projectile along the z-axis of the rotating frame 
which is taken to be parallel to $\vec R$. 
Eq.(\ref{RCC}) shows that only states with the 
same $\tilde K$ quantum number couple to each other. The dimension of 
the coupled-channels equations is thus considerably reduced. 

The third step is to solve the reduced coupled-channels equations 
for each  $\tilde K$=$-I_i,-I_i+1,....,I_i-1,I_i,I_i$ being the 
ground state spin of the Na projectile. Note that $K=I_i$ is fixed. 
As usually done for heavy-ion fusion reactions, 
we replace the proper boundary condition that the radial wave functions 
for the relative motion should be 
regular at the origin $R=0$ by the incoming wave boundary condition 
at an absorption radius $R_{abs}$ inside the potential pocket, and 
require 
\begin{eqnarray}
\tilde \Phi_{I\tilde K}^{JK}(R)\sim \left\{ \begin{array}{ll} 
T_{I\tilde K}^{L_iK}exp\Big[-i\int_{R_{abs}}^R(k_{L_iIK\tilde 
K}(R^\prime))dR^\prime \Big], & R\leq R_{abs}
\vspace{0.3cm}\\
H_J^-(k_{IK}R)\delta_{II_i}+
R_{I\tilde K}^{L_iK}H_J^+(k_{IK}R), & R\rightarrow \infty\\
\end{array} \right.
\label{ccbc}
\end{eqnarray}
with 
\begin{eqnarray}
\hskip -1cm & &k_{IK}=\sqrt{2\mu /\hbar^2(E-\epsilon_{IK})}
\label{wnasy}
\\
\hskip -1cm & &k_{L_iIK\tilde K}(R)=
\sqrt{\frac{2\mu}{\hbar^2}\Big(E-\epsilon_{IK}
-\frac{L_i(L_i+1)\hbar^2}{2\mu R^2}-V_0(R)-
V^\prime_{I\tilde K,I \tilde K }(R)\Big)}
\label{wnabsd}
\end{eqnarray}
In eq.(\ref{ccbc}), $T_{I\tilde K}^{L_iK}$ and $R_{I\tilde K}^{L_iK}$
are the transmission and reflection coefficients, respectively. 
We used these notations instead of $T_{I\tilde K}^{JK}$ and 
$R_{I\tilde K}^{JK}$ because all the transmission and the reflection 
coefficients are the same for a given $L_i$ irrespective of $J$.  
Once they are 
determined, the barrier transmission probability is given by 
\begin{eqnarray}
P^{J\tilde K}(E)=P^{(J)L_i\tilde K}(E)=\sum_{I}\frac{k_{L_iIK\tilde 
K}}{k}\Big|T_{I\tilde K}^{L_iK}\Big|^2
\label{pen}
\end{eqnarray}
where $k=k_{I_iK}$ is the wave number in the entrance channel. 
The first equality in eq.(\ref{pen}) 
means that the barrier transmission probability is 
the same for all $J$ for a given $L_i$.

The fusion cross section can be related to $P^{(J)L_i\tilde K}$ 
by calculating the flux at $R_{abs}$. To that end, we use the 
inverse transform of eq.(\ref{Twf}),
\begin{eqnarray}
\Phi_{I K L}^J (R)=\sqrt{\frac{2L+1}{2J+1}}\sum_{\tilde K}
(L0I\tilde K|J\tilde K)\tilde \Phi_{I \tilde K }^{JK}(R).
\label{iut}
\end{eqnarray}
Averaging over the initial and summing over the final spin projections, 
we obtain 
\begin{eqnarray}
\sigma_f(E)&=&\frac{\pi}{k^2}\frac{1}{2I_i+1}\sum_{\tilde K,L_i}
\sum_{J=L_i-I_i}^{L_i+I_i} 
(2J+1)P^{(J)L_i\tilde K}(E)
\label{crossd}\\
&=&\frac{\pi}{k^2}\sum_{\tilde K,L_i}
(2L_i+1)P^{L_i\tilde K}(E)
\label{cross}
\end{eqnarray}
We have used a simplified notation $P^{L_i\tilde K}(E)$ in 
eq.(\ref{cross}) 
because what one actually does is to solve the coupled equations 
eq.(\ref{RCC}) for each set of $\tilde K$ and $L_i$ and determine the 
barrier transmission probability. Eq.(\ref{cross})
reduces to the well known formula 
\begin{eqnarray*}
\sigma_f(E)=\frac{\pi}{k^2}\sum_{J}(2J+1)P^{J}(E)
\end{eqnarray*}   
for the fusion between two even-even nuclei 
by setting the initial intrinsic spin $I_i$ to be zero. 

\section{The charge and matter deformations of Na isotopes and their 
effects on the fusion reaction} 

We now discuss the charge and matter deformations of Na isotopes and 
the influence of their difference on fusion reactions taking 
$^{19,25,37}$Na+$^{208}$Pb scattering as examples. 
We perform the relativistic mean field (rmf) calculations with the NLSH 
parameters \cite{gam90,snr} to learn the r.m.s. 
radii and deformation parameters of 
the neutron, proton, charge and matter distributions of Na isotopes. 
Though the pairing correlation may strongly influence the 
accurate values of deformation 
\cite{yt97}, 
we ignore it, because our interest in this paper is to discuss 
global behaviors.
The results are shown in Fig.1, where 
the solid circles, solid squares and solid triangles represent 
the r.m.s. radii of the proton, matter and charge distributions, 
respectively. The open diamonds are the effective 
root mean square matter radii deduced from the measured interaction 
cross section using a Glauber-type calculation \cite{suzu98}.
We see that the theoretical results given by the solid squares well 
follow the general trend of the experimental data given by the open 
diamonds. 
Fig.1(b) shows the isotope variation of the quadrupole deformation 
parameters of the proton, matter and neutron distributions. 
It indicates that the charge and matter deformations are significantly 
different for many isotopes far away from the beta stability line. 

We now discuss how 
this difference influences the fusion reactions, or conversely whether 
one can probe this difference through the analysis of the fusion cross 
section. We choose the $^{19,25,37}$Na isotopes 
as representatives and analyze their fusion 
reactions with the $^{208}$Pb target.
As already mentioned, we treat $^{208}$Pb as a structureless particle. 

Before discussing the results of the coupled-channels calculations, 
we show in Fig.2 the potential barriers corresponding to fixed 
orientations of the deformed Na projectile 
assuming the quadrupole coupling, 
i.e. $\lambda=2$, 
\begin{eqnarray}
V(R,\theta)=\frac{-V_0}{1+exp[(R-R_0-R_P \beta^N_2 \sqrt{\frac{5}{4\pi}}
P_2(cos\theta))/a_0]}+\frac{Z_PZ_Te^2}{R}+
\beta^C_2 \sqrt{\frac{5}{4\pi}} f^C_2(R)P_2(cos\theta)
\label{potfor}
\end{eqnarray}
where the potential parameters are assumed 
to be $(V_0,a_0)=(114.2MeV,1.222fm)$. 
The radius parameters $R_P$ and $R_T$ 
have been calculated from the r.m.s. radius given by the 
rmf calculations according to 
$R_P(R_T)={\sqrt{\frac{5}{3}}}\times$ r.m.s. radius of 
the projectile(target). 
The higher and lower barriers in Fig.2 
are for 70.12 and 30.55 degrees, respectively. 
These are the effective angles in the so called orientation average 
formalism in the degenerate spectrum limit approximation 
when one discusses the effects of the ground state rotational band 
on the fusion of a deformed even-even nucleus by truncating at the 
first excited 2$^+$ state \cite{nbt}. 
The dotted line is the potential barrier when 
the effect of deformation is ignored. 
For each angle, the solid line is the potential barrier obtained by 
using different nuclear and Coulomb deformation parameters 
suggested by the relativistic mean field calculations. It is compared 
with the dashed line, which has been obtained 
by ignoring the difference between the charge and matter deformations. 
It is interesting to observe that the difference of the matter and 
charge deformations modify the barriers in opposite directions  
for proton and neutron rich isotopes.  

We now present the results of the coupled-channels calculations. 
There are no experimental data yet on the excited 
states of $^{19, 25, 37}$Na isotopes. 
As we stated before, we assume a strong coupling rotational model, where 
the K quantum number is assigned to be K=$\frac{3}{2}$ for all of them 
from the Nilsson diagram using the deformations suggested by the 
relativistic mean field calculations. 
The excitation energy is given by 
\begin{eqnarray}
\epsilon_{IK} =\frac{[I(I+1)-K^2]\hbar^2}{2\cal J}
\end{eqnarray}
where $\cal J$ is the moment of inertia. 

It is known that the experimental moment of inertia is smaller 
than the rigid body value ${\cal J}_{rig}$ by about a factor of 2 
\cite{bmII}.
Though the deviation of the moment of inertia from the 
rigid body value is somewhat smaller for odd mass nuclei \cite{bmII}, 
we assume,
\begin{eqnarray}
{\cal J} \approx {\cal J}_{rig}/2 
\label{choi1}
\end{eqnarray}
for the moment of inertia of $^{19,25,37}$Na nuclei. 
The rigid body value is given by 
\begin{eqnarray}
{\cal J}_{rig}=\frac{2}{5}AMR^2(1+\frac{1}{3}\delta)
\end{eqnarray}
where $\delta $ is the deformation parameter,
which is related to the quadrupole deformation parameter $\beta_2$ as
\begin{eqnarray}
\delta=0.945\beta_2[1-\frac{4}{3}\pi^2(a_0/R_P)^2]+0.34\beta_2^2.
\label{deltabeta}
\end{eqnarray}
In eq. (\ref{deltabeta}), the correction due to the surface diffuseness 
given through the diffuseness parameter $a_0$ 
has been included only to the leading order. 
The quadrupole deformation parameter $\beta_2$ is  
identified with the matter(nuclear) deformation parameter 
$\beta^N_2$ of the Na isotopes.
\par
We solve the coupled-channels equations taking up to the 
I$_{max}$=7/2 member of the ground state rotational band 
and including only quadrupole deformation 
for all three reactions.
The fusion cross section is then obtained by eq. (\ref{cross}). 
Once the fusion excitation function has been obtained, the fusion 
barrier 
distribution is calculated by the point difference formula of 
$\Delta E$=2MeV in the center of mass energy.
The results are shown in Fig.3. Each figure contains three theoretical 
lines. The dotted line has been obtained by the potential model. 
The solid line has been obtained by the coupled channels calculations 
using the charge and matter deformation 
parameters obtained from the relativistic mean field calculations. 
The dashed line has been obtained in the same way, but 
by ignoring the difference between the 
charge and matter deformations. 
We used the value of $\beta^N$ for both nuclear and charge 
deformations in this case.

These figures show that the effects of the difference between the charge 
and matter deformations are significant in the $^{37}$Na+$^{208}$Pb 
fusion reactions. They are noticeable also in the 
 $^{19}$Na+$^{208}$Pb fusion reactions.  
The existence of three or more peaks can be expected because of 
the truncation at 
I$_{max}$=7/2 by considering the degenerate spectrum limit. 
There appear indeed three separated peaks for the $^{37}$Na+$^{208}$Pb 
reaction. The shift of the highest and lowest 
peak positions by the 
difference between the charge and matter deformations accords with 
that shown in Fig.2 for the two channel problem.
Unfortunately, no visible effects are seen for the $^{25}$Na+$^{208}$Pb 
fusion reactions, which would be much more tractable experimentally 
than the other reactions.
Though $^{25}$Na has a large difference in the charge and matter 
deformations 
comparable to that in $^{19}$Na and $^{37}$Na, its effect on the 
fusion cross section is compensated by the counter effect of the 
difference between the charge and matter radii. 

The above conclusions might depend on the special choice of the 
moment of inertia given by eq.(\ref{choi1}).
In order to test this, we repeat the same calculations 
for $^{19,37}$Na+$^{208}$Pb reactions by 
changing the moment of inertia to 
\begin{eqnarray}
{\cal J} \approx {\cal J}_{rig}/3 
\label{choi2}
\end{eqnarray}
In Fig. 4, the dashed and the solid lines are the same as those 
in Fig.3 obtained by assuming eq. (\ref{choi1}). These lines 
are compared with the short dashed and the 
dash-dotted lines which have been obtained by assuming eq. (\ref{choi2}).
The figures show that the fusion barrier distribution 
distinguishes the different choice of the moment of inertia 
and that it clearly reflects the difference of the charge and matter 
deformations for both choices of the moment of inertia. 

\section{Summary}

The isotope variation of the charge and matter deformations is an 
interesting question, which will be addressed with the advent of 
secondary beam experiments. In this paper, we used relativistic mean 
field calculations to suggest that the charge and matter 
deformations can significantly differ from each other for some nuclei 
far away from the beta stability line. We then 
performed coupled-channels calculations using these informations, and 
have shown that the fusion excitation function and the fusion barrier 
distributions noticeably reflect the difference between the 
matter and the charge deformations in 
$^{19,37}$Na+$^{208}$Pb reactions. Unfortunately, 
this effect is compensated by the opposite effect of the difference 
between the matter and charge radii in the 
$^{25}$Na+$^{208}$Pb 
fusion reactions which will be more easily tractable experimentally. 

Finally, we wish to make some comments on the limitations 
of our relativistic mean field calculations. 
We ignored the pairing correlation in our 
relativistic mean field calculations of Na isotopes. 
The bulk properties of the light nuclei through helium to oxygen 
 has been calculated by the 
relativistic mean field theory 
with and without pairing correlation \cite{patra93}. 
The study shows that the 
r.m.s. radii of light nuclei are not so sensitive to pairing 
correlation, 
whereas the deformation parameters are affected considerably. 
We are now extending the relativistic mean field calculations to include 
the effects of pairing correlation in order to have a more reliable 
estimate of the deformation parameters. 

\section*{Acknowledgements}

This work is supported by the Grand-in-Aid for Scientific Research from 
Ministry of Education, Culture, Sports, Science and Technology 
under Grant No. 12047203 and No. 13640253. 


\newpage


\begin{center}
{\Large Figure Captions}
\end{center}

\noindent 
{\large FIG. 1}\\
The ground state properties of Na isotopes calculated by the 
RMF theory without pairing correlation. The top panel shows the proton, 
matter and charge r.m.s. radii. The open diamonds show the measured 
matter 
r.m.s. radius taken from \cite{suzu98}. The bottom panel shows the 
quadrupole 
deformation parameters of the proton, matter and neutron density 
distributions.

\noindent
{\large FIG. 2}\\
The effective fusion barriers for the s-wave scattering 
of $^{19,37}$Na from $^{208}$Pb as functions of 
the separation distance R between the projectile and target nuclei. 
The dotted line is the bare potential. The solid and dashed 
lines have been obtained by respecting the difference between 
the charge and matter deformations, and by ignoring it, respectively. 
For each system, the upper and the lower barriers are for the 
70.12 and 30.55 degree orientations, respectively.

\noindent 
{\large FIG. 3}\\
Effects of the difference between the nuclear and Coulomb deformations 
on 
(a) the fusion excitation functions and (b) the fusion 
barrier distributions in the $^{19,25,37}$Na+$^{208}$Pb reactions. 
The dotted lines are for the potential model.
The solid lines represent the results when the 
difference of the nuclear(matter) and the Coulomb(charge) deformations 
is taken into account. 
The dashed lines have been calculated by ignoring the difference of 
the charge and matter deformations.

\noindent 
{\large FIG. 4}\\
The sensitivity of 
(a) the fusion excitation function, and  (b) the fusion 
barrier distribution to the choice of the moment of inertia 
in the $^{19,37}$Na+$^{208}$Pb reactions. 
The dashed and the solid lines are the same as those in Fig.3.
They are compared with the short dashed and the 
dash-dotted lines obtained by assuming a different value for 
the moment of inertia in the coupled-channels calculations.

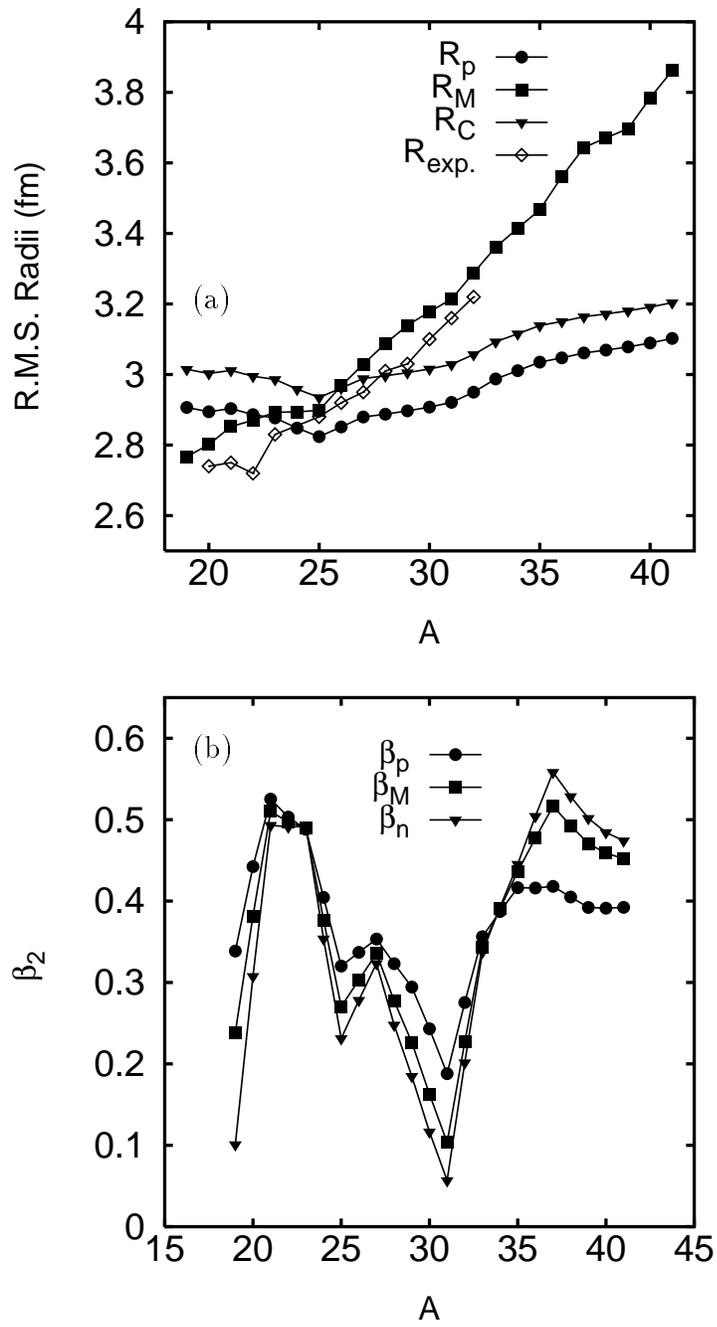
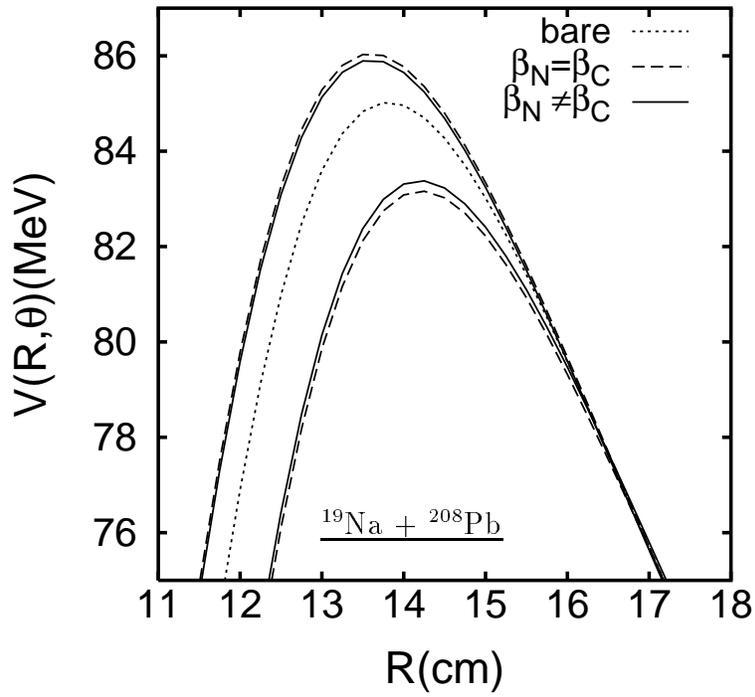
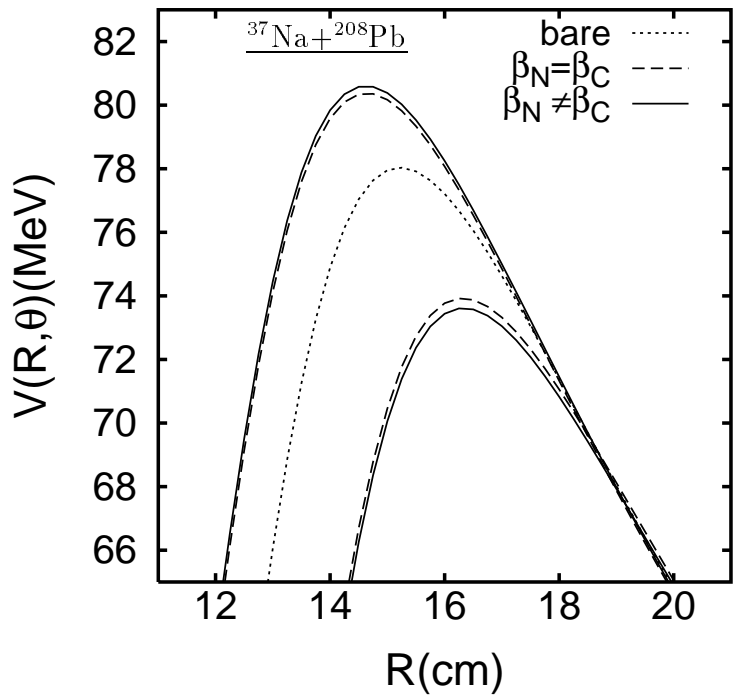
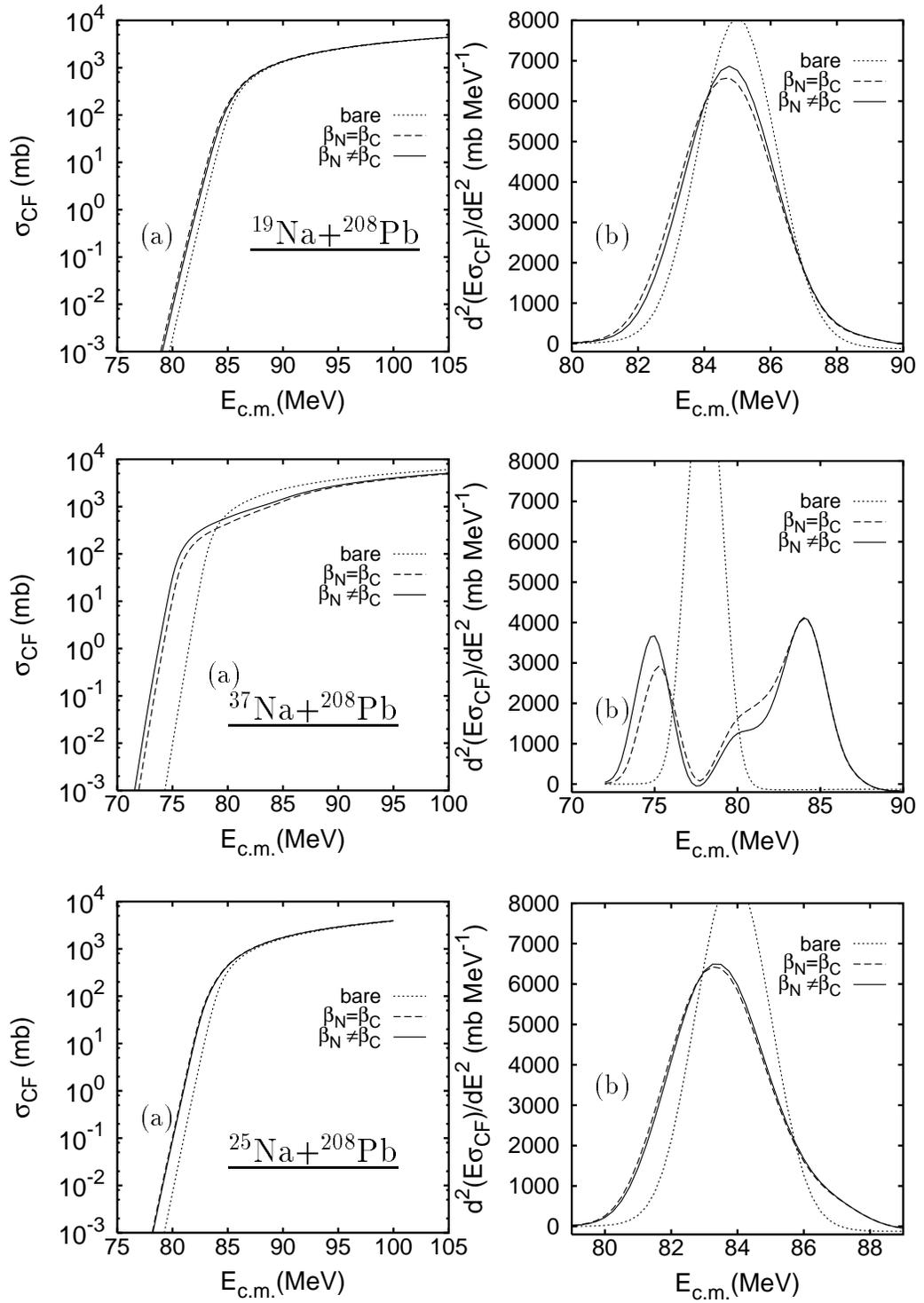
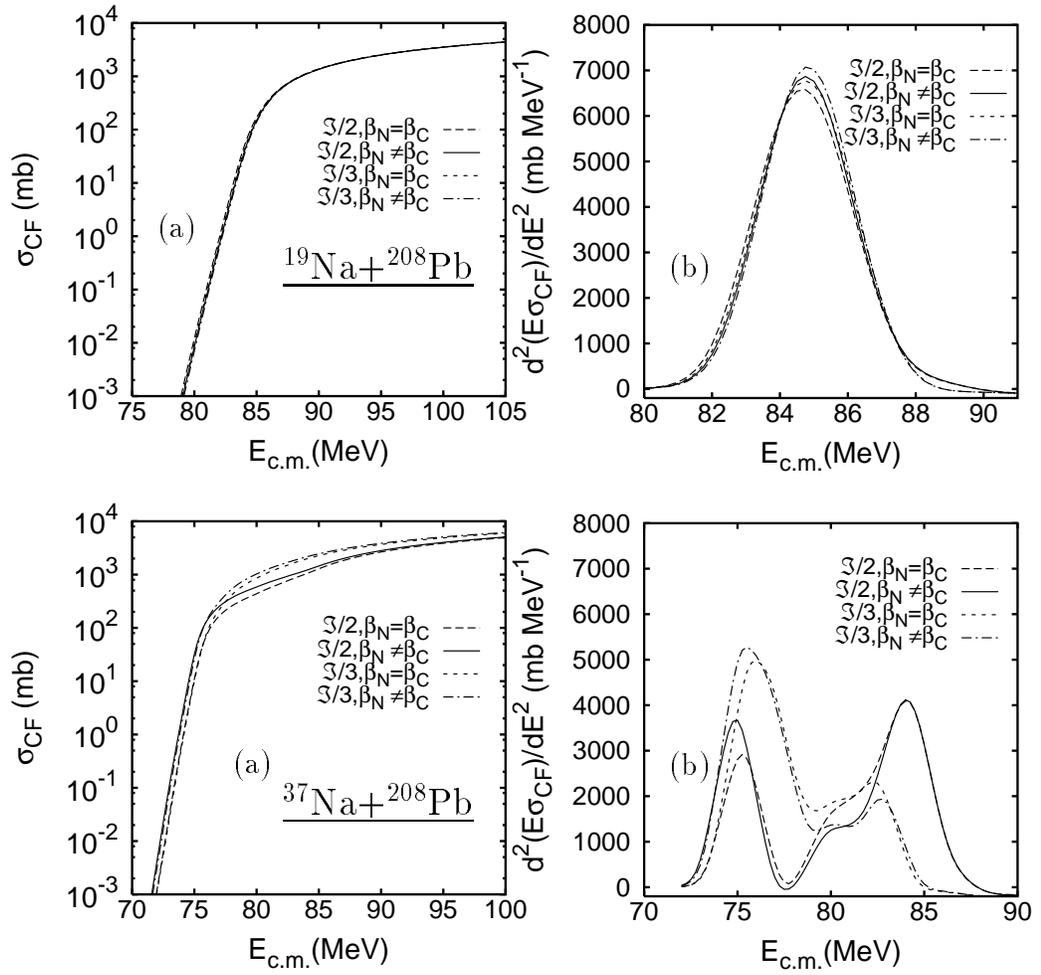
\end{document}